# A MACHINE LEARNING APPROACH FOR OPINION HOLDER EXTRACTION IN ARABIC LANGUAGE

Mohamed Elarnaoty, Samir AbdelRahman, and Aly Fahmy

Computer Science Department
Faculty of Computers and Information, Cairo University

m.elarnaoty, s.abdelrahman, and a.fahmy @fci-cu.edu.eg

## ABSTRACT

*Opinion mining aims at extracting useful subjective information from reliable amounts of text. Opinion mining holder recognition is a task that has not been considered yet in Arabic Language. This task essentially requires deep understanding of clauses structures. Unfortunately, the lack of a robust, publicly available, Arabic parser further complicates the research. This paper presents a leading research for the opinion holder extraction in Arabic news independent from any lexical parsers. We investigate constructing a comprehensive feature set to compensate the lack of parsing structural outcomes. The proposed feature set is tuned from English previous works coupled with our proposed semantic field and named entities features. Our feature analysis is based on Conditional Random Fields (CRF) and semi-supervised pattern recognition techniques. Different research models are evaluated via cross-validation experiments achieving 54.03 F-measure. We publicly release our own research outcome corpus and lexicon for opinion mining community to encourage further research.*

## Keywords

*Sentiment, Private state, Opinion mining, Conditional Random Fields, Cross-Validation.*

## 1. INTRODUCTION

With the increasing availability of blogs, internet forums and social networks, electronic press sites, people have the chance to express their opinions and sentiments and make them available to everyone. Opinion mining is concerned with the extraction and the computational treatment of such subjective information. The main tasks of opinion mining are:

1. The subjectivity extraction. This task aims at discriminating opinionated sentences from objective ones. Several researches explore this task on both document levels [1, 2, 3, 4, 5] and phrase level [2, 6, 7, 8, 9, 10, 11].

2. The opinion polarity identification. The objective of this task is to decide if a given subjective text has a positive or a negative or neutral orientation. Many publications have been dedicated to this task [2, 12, 13, 14, 15, 16, 17, 18]. This task can be considered the most active research point in opinion mining area.

3. Opinion elements identification task. The main components of any subjective expression are the opinion source, the opinion subject on which the opinion is expressed, and the opinion words or phrases. Few researches were published to address opinion components extraction task [19, 20, 21, 22, 23, 24].

4. The development of linguistic resources required for previous tasks such as subjectivity lexicons [25, 26, 27, 28, 29, 30] and annotated corpora [31].

The opinion holder extraction subtask gained less attention from opinion mining researchers than other tasks. To the best of our knowledge, there exists no published work in this task for Arabic language text, while for English language, we have few published works. One of the existing approaches to address this problem is to track the presence of subjectivity clues in order to identify opinionated uses of verbs, while using semantic parsing to locate opinion sources [19]. Another approach is to treat opinion holder extraction task as a sequential labelling classification problem [21, 23]. Structural features from syntactic parse tree are selected by other researchers to model structural relation between a holder and opinion expression [24]. The identification of opinion holders can benefit from, or perhaps even require, accounting for opinion expressions either simultaneously [19, 21, 23] or as a pre-processing step [22, 24].

Several challenges complicate the opinion source identification in Arabic language. Compared to the large number of publications and available resources and lexicons in English, the Arabic opinion mining field is still immature and has less number of publications [32, 33, 34, 35] and very few resources [36]. The lack of resources, the high inflectual nature of Arabic language [37], the variant sources of ambiguity [38], and rich metaphoric script usage remain the most challenging problems for Arabic NLP researchers. Opinion mining research, however, is affected by one more important limitation; the absence of a robust general purpose Arabic parser makes it difficult to understand the sentence structure and hence harden the extraction of the opinion holder as well as other opinion components.

In this paper, we explore the opinion source identification task in Arabic news text. Some researchers distinguish between subjective text documents, such as editorials and reviews, and objective documents, such as newspaper reports. We, on the other hand, do not make this distinction in our work. We believe that many newspaper articles contain a mix between both subjective and objective information and therefore, we need to identify sources of subjective information inside these articles. We explore three different approaches to solve Arabic opinion holder extraction problem. The three investigated approaches are semi-supervised traditional pattern based classification, supervised machine learning sequential labelling (CRF), and an integration between them.

This paper contribution is the following: First, it is the first Arabic research work to address opinion holder extraction problem. Second, two new features for this task are introduced: 1) the semantic field feature as a mighty indication for subjectivity existence, 2) the named entity feature as opinion holders are often identified as a named entities. Finally, we publically released an annotated Arabic corpus for opinion holder and an Arabic subjectivity lexicon resource for Arabic NLP researchers.

The paper is organized as follows: Section 2 gives a general overview of opinion source identification task and mentions related works. Section 3 presents the proposed solution and its main components. Section 4 focuses on opinion holder extraction using patterns while section 5 focuses on the CRF tagging and used features. Section 6 shows the experimental results and error analysis. Finally, Section 7 draws our conclusions and future work.

## 2. TASK DESCRIPTION AND RELATED WORK

There are three main ways that private states are expressed in language: speech events expressing private states, direct mentions of private states, and expressive subjective elements (Banfield, 1982) [39]. In this work, we distinguish between three types of opinion holders corresponding to the above three types of subjective statements:

| | |
|---|---|
| **Direct Speech Event Holder Example** | • و قال رئيس الجمهورية الأسبق " الوزيرة الأمريكية طرحت مجرد أفكار و لم تقدم حلا عمليا "<br>• And the former president said "The American minister posed some ideas but didn't present a practical solution" |
| **Indirect Speech Event Holder Example** | • و قد أكدوا أن الرئيس قد قال إنه سعيد بالعمل<br>• They did ensure the president said he is pleased with the work |
| **Private State Expression Holder Example** | • و أيد الرئيس اقتراح وزير التعليم<br>• And the president supported the Education Minister's suggestion |
| **Expressive Subjective Elements Holder Example** | • تزاحم الجمهور حول اللاعب للحصول على توقيعه<br>• The crowd has gathered around the player to obtain his signature |

Figure 1. Opinion Holders Three Types Examples

1- Opinion Holders For Speech Events Expressions, Either Direct Or Indirect.

By speech events we mean that some subjective statement is said by someone directly or claimed to be said by him. The first case is referred to as direct speech event while the second case is indirect speech event.

Considering the examples presented in Figure 1, the first sentence is an example of direct speech event and the opinion holder in this sentence is "the former president". The second sentence (Figure 1) is an example of an indirect speech event with the word "president" as an opinion source. Opinion holders for direct/indirect subjective speech events are referred to along this paper with *"Type 1 Holders"*.

2- Opinion Holders For Private State Expressions.

By private state expressions, we mean the use of some key subjective words, mostly verbs, which express certain sentiments about opinion subjects. Words such as {liked, hated, supported, was angry with …etc} are used to express someone's feeling about something.

For instance, the use of verb "supported" in the third sentence (Figure 1) causes us to nominate the word "president" as the opinion source of this subjective sentence. Opinion holders for this type of subjective statements are referred to along this paper with *"Type 2 Holders"*.

3- Expressive Subjective Elements Sources.

With expressive subjective elements, sarcasm, emotion, evaluation, etc. are expressed implicitly through the way something is described or through particular wording. Clearly, this type of subjective expressions is the most difficult one to detect, as the subjectivity analysis of the statement in this case depends on understanding its meaning rather than its structure.
For example, while there are neither any sentiment words nor any speech events in the last sentence (Figure 1), it's obvious that the sentence expresses the crowd admiration sentiment towards their favorite player. Opinion holders for this type of subjective statements are referred to along this paper with *"Type 3 Holders"*.

The proposed work aims at identifying all the three types of the opinion holder within Arabic news based on pattern matching and sequential labelling.

Choi et al [21] presented an approach that combines conditional random fields and extraction patterns. They treat opinion source finding as a combined sequential tagging and information extraction task. They exploit the high precision but low recall extraction patterns as a feature to train CRF engine along with other certain collection of lexical, syntactic, and semantic features. CRF as well was used to detect opinion expressions. They achieve the opinion source identification task with 62.0 F-measure points.

Choi et al [23] use CRF sequential tagging classifier to extract n-best candidates for direct expressions of opinions and their sources jointly, and employ integer linear programming to find the link between subjective expressions and their sources jointly with 69.0 F-measure success.

## 3. THE PROPOSED APPROACH

In this section we start first by describing the needed pre-processing phase to the input Arabic text (Subsection 3.1), and then provide a short overview of the investigated approaches (Subsection 3.2). The details of these approaches are given in sections 4 and 5 and the results analysis is given in section 6.

### 3.1 Corpus Collection and Preparation

Arabic NLP resources are rare, therefore we had to build the needed resources, and perform the required pre-processing on it. The pre-processing tasks include sentence segmentation, morphological analysis, part of speech tagging (POST), semantic analysis, named entities recognition (NER), subjective analysis, and manual annotation for opinion holders.

Arabic search engine (http://www.alzoa.com/) was used to crawl the web for Arabic news articles getting 150 MB news documents. 1 MB only of our corpus was manually tagged for opinion holder by three different persons, where conflicts of tagging are resolved using majority voting principle. The tagged corpus is freely distributed by the Arabic Language Technology Center "ALTEC" by following the data resources link on their web site (http://altec-center.org/)

We used the Research and Development International (RDI) (http://www.rdi-eg.com/) toolkit to handle the orthographic and morphological analysis of Arabic sentences, part of speech (POS) tagging, and semantic analysis of news words.

The Research and Development International (RDI) toolkit mainly consists of Arabic RDI-ArabMorpho-POS tagger [40] and RDI-ArabSemanticDB tool [41]. RDI-ArabMorpho-POS tagger includes Arabic morphology and POS models. The tagger works with an average of 90.4% accuracy. RDI-ArabSemanticDB tool is composed of Arabic lexical semantics language resource (database) and its related interface. The database archives approximately 40,000 Arabic words, 1840 semantic fields, and 20 semantic relations, such as synonyms, antonym, hyponymy and causality.

For Arabic named entity recognition (ANER), we employed the bootstrapping technique described by S.AbdelRahman et al [37]. The named entity tagger recognizes named entities for ten named entity (NE) classes, {Person, Location, Organization, Job, Device, Car, Cell Phone, Currency, Date, and Time}, with F-measure accuracies of {74.06%, 89.09%, 75.01%, 69.47%, 77.52%, 80.95%, 80.63%, 98.52%, 76.99%, and 96.05%}, respectively.

For subjectivity classification of sentences, we used two naive classifiers that were described by E.Riloff and J.Wiebe [6]. The first classifier looks for sentences that can be labeled as subjective with high confidence based on the existence of certain subjectivity clues. The second classifier looks for sentences that can be labeled as objective with high confidence. All other sentences in the corpus are left unlabeled.

The subjectivity clues are divided into strongly subjective and weakly subjective clues. A strongly subjective clue as defined by E.Riloff and J.Wiebe [6] is the one that is seldom used without a subjective meaning, whereas a weakly subjective clue is the one that commonly has both subjective and objective uses. The high-precision objective classifier classifies a sentence as objective if there are no strongly subjective clues and at most one weakly subjective clue in the current, previous, and next sentence combined while the high-precision subjective classifier classifies a sentence as subjective if it contains two or more of the strongly subjective clues [6].

For obtaining Arabic strongly and weakly subjectivity clues, we manually translated the MPQA subjectivity lexicon developed by Pittsburgh University (http://www.cs.pitt.edu/mpqa/subj_lexicon.html) [25] into Arabic and marked the polarity and strength for each word. The translated MPQA subjectivity lexicon contains more than 8000 English words and corresponding Arabic words and it is made available through ALTEC Society (http://altec-center.org/)

As in English lexicon, words that are subjective in most contexts were marked strongly subjective (*strongsubj*), and those that may only have certain subjective usages were marked weakly subjective (*weaksubj*). For example, the word "condemning - مستنكرا" is a *strongsubj* clue while the word "respectful - محترم" despite its, full of praise, meaning is a *weaksubj* clue due to its frequent use in formal letters and reports.

## 3.2 The Investigated Approaches

For opinion source identification, we explored three different approaches.

1. Opinion sources identification using traditional pattern matching.

2. Opinion sources identification using sequential tagging CRF classifier.

3. Opinion sources identification using sequential tagging CRF classifier with the use of patterns as a feature

The first investigated approach made use of hand crafted holder patterns. Such patterns are defined in terms of words, named entities and POS tags. The pattern matching approach is described in details in section 4.

The second investigated approach is supervised statistical machine learning approach. We used conditional random field classifier to train our model using our feature set. The CRF features are described in details in section 5,

The third approach is also based on CRF. The only difference between the second and third approaches is the matched pattern feature being disabled in the second approach and enabled in the third approach.

## 4. OPINION HOLDER EXTRACTION USING PATTERN MATCHING

An initial set of hand crafted patterns were extracted by manual inspection of a collection of articles crawled from the web. The patterns are defined using key phrases, and POS tags (Figure 2).

To test the manually extracted patterns validity; two tests are done. The first test is the frequency test where the pattern is retained if its occurrence frequency in a corpus of 150 MB text exceeds a threshold value (the value 5 is selected by trial and error method).

The second test is the precision of the pattern in extracting the opinion holders from a corpus of size 1 MB. All patterns of precision lower than a selected threshold were removed. The selection of the threshold is subject to a tradeoff between recall and precision. The larger the value of the threshold, the less patterns we extract, and hence the less recall we obtain, while the smaller the value of the threshold, the less precision we obtain. Experimentally, 0.8 was selected as the threshold value.

The previous process resulted in a total of 43 patterns neglecting the morphological inflection of words ("he said قال" and "she said قالت" are counted in the same pattern); examples of the final patterns set are shown in Figure 2.

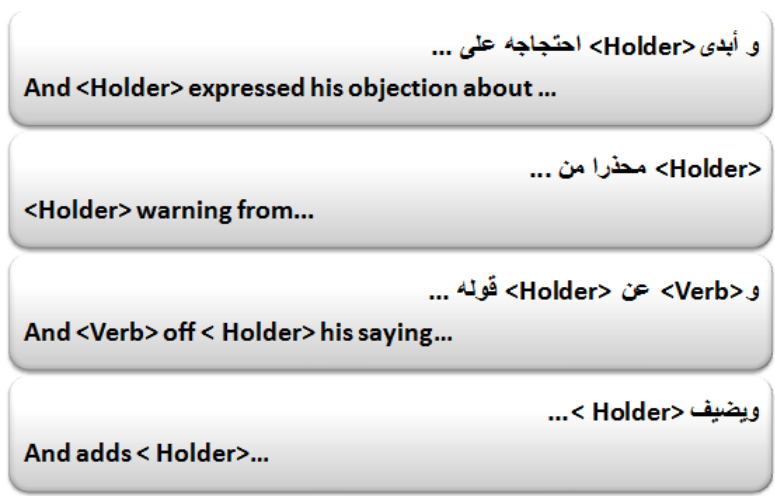

Figure 2. Opinion Holder Patterns Examples

Finally, our pattern based opinion holder classifier was run to identify the opinion sources on the testing data. The classifier used the patterns as input for the pattern matcher code and all matched candidates were classified as following:

1.  The candidate opinion holder is neglected if its containing statement is of type objective (as determined by the high-precision objective classifier).

2.  The candidate opinion holder is retained if its containing statement is of type subjective (as determined by the high-precision subjective classifier).

3.  The candidate opinion holder is also retained if it is a named entity and its containing statement is neither objective nor subjective from the point of view of the high precision classifiers.

Section 6 will give examples of correctly and erroneously identified opinion holders, and provide the evaluation for pattern classifier results.

## 5. OPINION HOLDER EXTRACTION USING CRF

The opinion holder extraction task can be formulated as a classification problem for each word in the corpus, where each word is classified as one of three categories {B-Holder, I-Holder, or Non-Holder} based on the word features. We used for this classification problem the CRF++ (http://crfpp.sourceforge.net/) classifier.

CRF is a discriminative probabilistic model [42]. It is used for segmenting and labelling the sequential data. It is a generalization of Hidden Markov Model in which its undirected graph consists of nodes to represent the label sequence $y$ corresponding to the sequence $x$. The aim of CRF model is to find $y$ that maximizes $p(y \mid x)$ (Equation 1) for that sequence.

$$p(y \mid x) = \frac{1}{z(x)} * \exp(\sum_{t} \sum_{k} \lambda_k f_k(y_{t-1}, y_t, x))$$

$$z(x) = \sum_{y \in Y} \exp(\sum_{t} \sum_{k} \lambda_k f_k(y_{t-1}, y_t, x)) \qquad (1)$$

$$; \lambda_k \text{ is the weight of } f_k$$

A CRF model was trained on the manually tagged for opinion holder training corpus using a set of features extracted from text. Converting a piece of text into a feature vector representation is an important part of data-driven approaches to text processing. There is an extensive body of work that addresses the selection of the most salient features for machine learning classifiers in general, as well as for learning approaches tailored to the specific problems of classic text categorization and information extraction [43, 44, 45].

We compiled a set of morphological, lexical, and semantic features to train our CRF classifier. Unlike Choi et al work [21], we didn't use any syntactic parser features due to lack of robust general purpose Arabic parsers. On the other hand, we exploited some features that are not used by Choi et al such as named entities and semantic field features.

We used two classes of features for training the opinion holder CRF model:

1. Window features: We used a window of size 2n+1 around the current word. Most of features fall in this type. We will refer to window features in the following by using the suffix letter *w*.

2. Word features: They are used without considering surrounding context of the current word.

The full feature set according to this categorization is described below:

**1. The Word And Its Surrounding (*w*):**

The word by itself was used as a feature and also the previous and next 3 words. The logic behind this is obvious as we need to train the classifier for common opinion holder surrounding words which are considered as keywords.

**2. The Semantic Field (SF) Feature (w):**

The occurrence of some keywords may be sufficient for humans to detect an opinion source, for example, "clarified وضّح" keyword is an obvious clue for opinion holder existence in case the statement was tagged as subjective statement. But what if the training set contains small or even zero frequency of this keyword? We can compensate this by grouping together the semantically related keywords like (illustrated, showed, made clear...etc) (بيّن, أبدى...إلخ) so that the missed keyword ("clarified" in our example) has the same conditional probability *p(OH|SF)* of its synonyms.

This feature was determined using the RDI semantic lexicon [41], the input to the lexicon is the word in its phrase and the output is the semantic field of this word. If the word doesn't exist in the lexicon, it's given a null value for this feature. Prepositions also have no semantic fields and hence, take null value.

**3. Part Of Speech Tag (POST) Feature (w):**

Part-of-speech (POS) information is commonly exploited in sentiment analysis and opinion mining. One simple reason holds for general textual analysis, not just opinion mining: part-of-speech tagging can be considered to be a crude form of word sense disambiguation [46].

The word POST feature is extracted using the RDI morphological analyzer [40]. The actual POST tag from RDI was not used as it's so extensive that the same tag may not occur frequently, Figure 3 shows an example of the RDI actual tagging result:

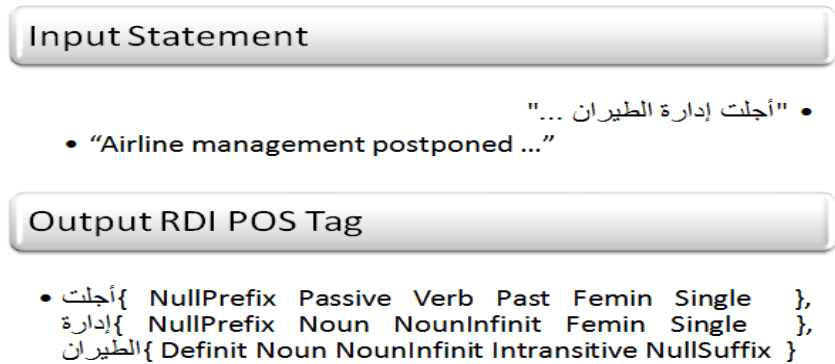

Figure 3. RDI Part Of Speech Tagging Example

Hence, we reduced this tag to a smaller tag set {Noun, Definite, Verb, Translit, Number, Symbol, and NA}. Any non-Arabic words (Latin letters, special characters...etc) are tagged as symbols. Definite class is any noun that is attached with a definite article prefix (ال). Translit class is dedicated for transliterated words. The RDI POS tagging in Figure 3 will be reduced to:

<div dir="rtl">{ Definit الطيران }, { Noun إدارة }, { Verb أجلت }</div>

Opinion holders could be nouns, definites, or translits. But the other tag set classes can be useful for recognizing the opinion holder context. For example, an opinion holder in direct speech events subjective statements are usually preceded by a word with verb POS tag.

## 4. Base Phrase Chunk (BPC) Feature (w):

PBC represents the Base Phrase Chunks (atomic parts) of a sentence. The BPC feature is useful especially for detecting the boundaries of the recognized opinion sources.
This feature was calculated from Yamcha [47] training toolkit (http://chasen.org/~taku/software/yamcha/) using word and POST features. Possible BPC tags are {B-Tag, I-Tag, O} | Tag $\in$ {NP,VP,PP,CONJP,ADJP,ADVP}

## 5. The Named Entity Features (w):

NE features are a set of Boolean features indicating whether each word in the moving window is tagged as a named entity or not. Each NE class has a corresponding Boolean feature column. These feature were calculated by applying bootstrapping technique described by S.AbdelRahman et al [37] on our corpus to detect NE's of 10 different classes; Person, Location, Organization, Job, Device, Car, Cell Phone, Currency, Date, and Time.

Of these 10 mentioned classes, person, and job NE's are the most effective NE features in the detection of opinion holders as the vast majority of opinion sources are either persons, or jobs.

Another advantage of using this feature is that the boundaries of opinion holder are correctly detected following the detection of corresponding NE boundaries. We were able to benefit from this characteristic in evaluating opinion holder extraction based on exact matching scheme.

**6.  Pattern Feature :**

It is a Boolean feature that indicates whether the current word is a part of any of the opinion holder patterns described in section 4 or not. We made two experiments, one of them with this feature enabled and the other with this feature disabled to figure out the effect of adding pattern as a feature to the ML CRF training as will be discussed later (Section 6).

**7.   MPQA Subjectivity Lexicon Features:**

MPQA features are four binary features that were extracted from the Arabic version of MPQA subjectivity lexicon.

7.1 Strong Subjectivity Clue Feature (w):

   It is a binary feature that indicates whether the word is a strong clue for subjectivity or not as retrieved from MPQA lexicon.

7.2 Weak Subjectivity Clue Feature (w):

   It is a binary feature that indicates whether the word is a weak clue for subjectivity or not as retrieved from MPQA lexicon.

7.3 Subjectivity Classifier Feature :

   It is a binary feature that indicates whether the current word is a part of a statement that was classified as a subjective statement using the high-precision subjective classifier.

7.4 Objectivity Classifier Feature :

   It is a binary feature that indicates whether the current word is a part of a statement that was classified as an objective statement using the high-precision objective classifier.

Section 6 will give examples of correctly and erroneously identified opinion holders, and provide the evaluation for CRF classifier results.

# 6. EXPERIMENTAL ANALYSIS

The evaluation of the three approaches described previously is based on the exact match between identified opinion holders and the manually annotated true holders. We used precision, recall and F-measure to evaluate these approaches

$$precision \quad = \frac{True \quad Positive \quad H\text{ olders}}{retrieved \quad H\text{ olders}} \qquad (2)$$

$$recall \quad = \frac{True \quad Positive \quad H\text{ olders}}{relevant \quad H\text{ olders}} \qquad (3)$$

$$F\text{ - }measure \quad = 2 * \frac{(\ precision \quad * \ recall \quad )}{(\ precision \quad + \ recall \quad )} \qquad (4)$$

In the patterns experiment, we applied pattern matching on the entire corpus. In CRF experiments, since we have only 1 Mb annotated text and in order to avoid the problem of over-fitting to training set, we applied 3-fold cross-validation method to verify the achieved results.

Table 1 shows the results for our three investigated approaches. CRF machine learning technique results are better than pattern matching ones in terms of recall and precision, and F-measure. The integration between patterns and CRF improves the F-measure for all three CRF folds experiments. Figure 4 shows the comparison between pattern results and CRF three folds results.

In the CRF experiments, the CRF third fold achieves the best results in terms of F-measure. Fold 2 achieves comparable results; however, fold 1 has about 10.0 points less than the other two folds. The average F-measure of the three folds is 49.22 without using pattern feature, and 50.52 using pattern feature.

Table 1. Opinion Holder Extraction Results

| Technique | Dataset | Precision | Recall | F-measure |
|---|---|---|---|---|
| Pattern results | - | 29.93 | 30.44 | 30.18 |
| CRF results | Fold1 | 66.67 | 29.74 | 41.13 |
| | Fold2 | 84.14 | 38.73 | 53.04 |
| | Fold3 | 84.83 | 39.05 | 53.48 |
| Integration results | Fold1 | 70.45 | 31.79 | 43.82 |
| | Fold2 | 86 | 39.05 | 53.71 |
| | Fold3 | **85.52** | **39.49** | **54.03** |

Returning to the work of Choi et al [21], we found that the behaviour of the pattern experiments compared to the ML experiments is similar in both English and Arabic language. Choi et al [21] raise the recall of tagged holders from 41.9 in case of patterns experiment to 51.7 in case of CRF experiment, and raise the precision from 70.2 to 72.4 resulting in increasing F-measure from 52.5 to 60.3. The combination of the two approaches gives 54.1, 72.7, and 62.0 recall, precision, and F-measure, respectively, which means that encoding the pattern as a CRF feature enhances the English opinion holder extraction task with 1.7 F-measure points. All previous results are reported based on exact matching method.

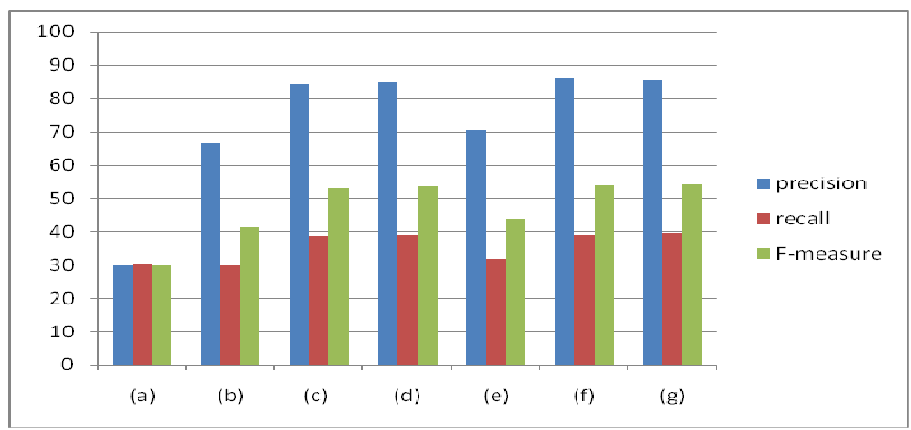

Figure 4. Opinion Holder Results: a) Pattern Only, b) Fold1 CRF, c) Fold2 CRF, d) Fold3 CRF, e) Fold1 Integration, f) Fold2 Integration, g) Fold3 Integration

## 6.1 Discussion:

From our experiments, CRF proves that it is more practical in opinion holder recognition problem than patterns. This is conformed to the work of Choi et al [21] conclusion. CRF sentiment training is able to capture most instances of patterns using our proposed feature set, and it is able to capture other types of holders that patterns fails to cover. Most of CRF failure cases are due to sentiment expression complexity and language inherent ambiguity which also restrain the pattern matching task.

Another drawback for pattern work is that patterns are applied only to sentences that are not tagged as objective. Therefore, many holders are lost due to erroneous tagging of their statements as objective. This is not the case of CRF, as the sentence subjectivity is encoded as a feature for CRF model training, and whatever the conditional probability of being a holder given existence in a subjective statement, there is still a probability for being a holder in a statement that is tagged as objective.

With no more true holders captured by pattern, and with low precision nature of patterns, it is justified why CRF outperforms patterns in terms of precision, recall, and F-measure. Although the pattern is added as a feature to our CRF experiment, its effect is minor compared to the other features. After we disabled the pattern feature, the average F-measure is just decreased by 1.3 F-measure points from 50.52 to 49.22. Choi et al [21], on the other hand, get 1.7 F-measure enhancement from adding the pattern feature in English.

CRF technique also is successful in detecting the boundaries of opinion holders. CRF utilizes BPC (base phrase chunking) feature besides the named entities boundaries to precisely recognize holder boundaries. While semantic field feature is used for detecting sentiment words (verbs, adjectives, and adverbs) which ease the task of capturing private state expressions. MPQA features are utilized by CRF for detecting the context of the holder as well as deciding the subjectivity of the sentence.

Table 2 shows the effect of pattern, NE's, MPQA, and semantic field features. The first row presents the F-measure for opinion holder extraction on fold 3 dataset using all features. Following rows show the effect of the absence of studied feature on the F-measure.

Table 2. Comparing CRF Features' Contributions To Final Result

| Disabled Feature | F-measure |
|---|---|
| None | 54.03 |
| Pattern feature | 53.48 |
| Subjectivity classifiers features | 52.85 |
| Semantic Field feature | 51.52 |
| Subjectivity clues features | 51.21 |
| Job NE only | 40.19 |
| Person NE only | 34.52 |
| All NE features | 23.76 |

Table 2 shows that NE features absence degrades the performance severely. The absence of person only NE decreases the F-measure approximately by 20 points which indicates how important is this feature in the recognition of opinion sources. MPQA and Semantic field features contributes significantly to the enhancement of the result in the range of 1.2 to 3 F-measure points.

To explain the performance of CRF classifier in recognizing different types of opinion holder, we consider our fold 3 CRF experiment. Table 3 shows that the opinion holders in fold 3 dataset are fairly distributed among the 3-type opinion holder. However, for correctly extracted opinion holders, type 1 covers more than 55% of the extracted true holders. Holder types 1 and 2 cover together around 88% of the extracted true holders. The main difficulty arises from the third type of holders due to the related complicated and unusual sentence structure, and its dependence on deep semantic knowledge. Pattern recognition, on the other hand, fails to detect most of types 2 and 3 holders, while it performs moderately in detecting type 1 holder.

Table 3. Opinion Holder Three Types Results Comparison

| Holder Type | Type 1 | Type 2 | Type 3 |
|---|---|---|---|
| percentage | 37% | 38% | 25% |
| detected holders ratio | 55.65% | 32.26% | 12.09% |
| Accumulative ratio | 55.65% | 87.91% | 100% |
| Type recall | 66.35% | 35.09% | 15.78% |
| Accumulative recall | 66.35% | 50% | 39.49% |

The number of captured opinion holders of a certain type relative to the total number of true holders of the same type defines this holder type extraction recall. Table 3 shows the recall for all three opinion holder types. Type 3 achieves over than 65% recall. If we calculated the accumulative recall for types 1 and 2, we still get 50% recall (Table 3, and Figure 5), while the overall recall for all three types is degraded lower than 40% (Tables 1, and 3).

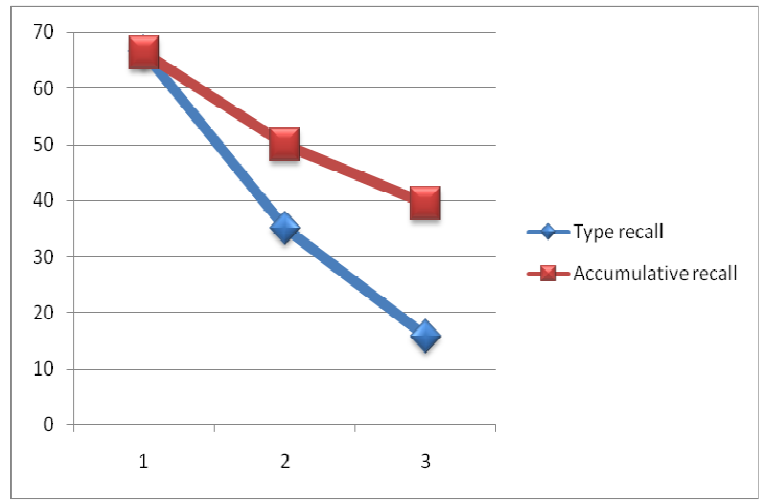

Figure 5. Opinion Holder Types Mining Recall

The reason that CRF fold 1 experiment results are less than the other two folds is that fold 1 testing dataset contains relatively small number of holders (195); most of them are of holder types 2 and 3. On the other hand, the opinion sources in folds 2 and 3 are approximately uniformly distributed on three opinion holder classes. Therefore, these parts achieve better results but still incomparable to the sentiment source identification results in English language.

From Natural Language Processing point of view, the quality of the opinion holder extraction task relies on the performance of several pre-required NLP tasks such as part of speech tagging, base phrase chunking, named entity recognition, and sentiment analysis. While all these pre-tasks are useful in the detecting the holder, these tasks in the same time adds an incremental error ratio to the opinion holder extraction task. If we add to this factor the high inflectual nature and the relative complexity of Arabic language sentence structures compared to the Latin languages, we can explain the degradation of this task results compared to the other languages like English. Generally speaking, the performance degradation can be returned mainly to the following two factors:

1. The incremental error in pre-required tasks in English is smaller than its peer incremental errors in Arabic Language NLP tasks.

2. The use of lexical parser features in other languages, while we have not used any parsers yet. The robust, publicly available, Arabic parser existence is a questionable research compared to English existing mature parsers.

## 6.2 The Error Analysis:

Starting with CRF experiments results, we show first some examples of correctly tagged statements. Afterwards, false positives and false negatives examples are presented. In all subsequent statements, we underline all NPs tagged by the classifiers as opinion sources while the non-detected sources remain only bold without underlining. It's worth mentioning, as shown in the examples, that we include the titles, surnames, nationalities and job titles in the tagged opinion sources for both training and testing corpora. This is useful as some the opinion sources are mentioned by job title instead of person name. For instance, "Algerian President الرئيس الجزائري" may be mentioned in text without using his name.

**Correctly Detected Holder Example:**

1- وقال **سى كى شو مدير الشركه بالسعوديه**ان الجهازين حازا العديد من الجوائز العالميه لحجمهما الانيق وادائهما الرائع

And **Si Kee Shu the company CEO in Saudi Arabia** said that both devices achieved many international prizes for their elegant size and wonderful performance.

2- ووصلت رايس اسرائيل قادمه من بيروت حيث عقدت اجتماعا مع **رئيس الحكومه اللبنانيه فؤاد السنيوره** الذى شدد على ضروره التوصل الى وقف فورى لاطلاق النار

And Rice arrived at Israel coming from Beirut where she hold a meeting with **Lebanon Government President Faud Al-Sanyoura** who emphasized the necessity of immediate stop of fire.

3- وقد احب **اينشتاين** طرائق التدريس فيه

And **Einstein** liked the ways of teaching there

4 – ولم يكن **اينشتاين** من دعاه الحرب

And **Einstein** was not from "callers for war" party

The first two statements are two examples of correctly tagged holders in direct speech events. The third statement is an example of private state expression holder that is correctly tagged. The last statement can be considered an example of a sentiment holder of the third type as it expresses Einstein feelings against wars by mentioning his negative actions towards its declaration.

**False Positives in CRF Experiments:**

1-    وظهر <u>**اللاعب السنغالي سيدي بيه**</u> بمستوى عادى

And <u>**the Senegalese player "Sedi Beah"**</u> appeared in average level

2-    لم يتمكن <u>راهب الفكر</u> من الصمود

<u>**Monastic of thought**</u> couldn't resist

3-    كما دعا <u>**الشعب المصري**</u> الى التفكير فى قضيه التبرع بالقرنيه

And called out <u>**Egyptian people**</u> to think about Cornea donation subject

In the first two examples, both sentences are definitely subjective statements. This could be easily figured out from the use of words like (average عادي, monastic of thought راهب الفكر), but the opinion holders here are the writers of the two sentences and not the two subjects in each sentence.

The third sentence is an example of inherent ambiguity of Arabic language. Even with the existence of lexical parser, it will be difficult to know whether the (Egyptian people الشعب المصري) is the subject or object of the verb (called out دعا), we know from the previous context that Egyptian people here is the object of the verb and the caller is some doctor who motivates others to donate with corneas. Tagging the previous statement as subjective is also a controversial manner. We consider it subjective as not all people encourage the donation of their corneas even after their death.

**CRF Experiment False Negatives:**

**1-**    عرف عن عصفور الشرق <u>**توفيق الحكيم**</u> انه كان اقل عداء للمراه من <u>**العقاد**</u>

East Sparrow **Tawfik Al-Hakim** was known to be less aggressive to woman than **Al-Akkad**

2-    ولم يكن غريبا ما حظيت به الموديلات الجديده مثل مورانو و ارمادا و الباثفيندر الجديده من اعجاب واهتمام خاص من قبل <u>**المواطنين**</u>

And it wasn't strange what new models such as Murano, Armada, new Pathfinder got from **citizens**, both admiration and interest

3-    واستغل <u>**اينشتاين الابن**</u> الفرصه السانحه للانسحاب من المدرسه فى ميونخ التى كره فيها النظام الصارم والروح الخانقه

And **Einstein the Son** made use of this chance to leave the school in Munich, where (he) hated its strict regime and choky spirit

4-    خرج <u>**العرب والمسلمون**</u> بمئات الالاف ومن اللحظه الاولى لقصف المدن والقرى اللبنانيه للتضامن مع الحزب ولبنان والاعراب عن الغضب تجاه العدوان الذى يتعرضان له

**Arab and Muslim people** went out in streets in huge numbers (hundreds of thousands) from the first moment of attacking Lebanon cities and villages to express their support for party and Lebanon and their anger for the attack they face

5-    تظاهر <u>**عشرات من الباكستانيين**</u> امس فى مدينه كراتشى ضد اى حرب محتمله تقودها الولايات المتحده ضد العراق

**Tens of Pakistanis** protested in Karachi against any possible war towards Iraq leaded by USA

The first two sentences are examples of common failure to detect type 2 holders due to the sentence structure complexity. In the following statement, the classifier couldn't recognize that the verb "hated كره" refers to previously mentioned Einstein. This is an example of a case where the holder classification is much simpler for English than Arabic because the personal pronoun "he" is used in English translation while it is implicit in the Arabic statement.

The last two statements are examples of common failure to capture type 3 holders, as the sentiment of (Arab and Muslim people العرب و المسلمون, Pakistanis الباكستانيين) is not directly stated in the two statements but could be grasped from the actions of demonstration and protest.

**False positive And False Negative in The Same Sentence:**

- وقالت **الرئاسه** شدد **الرئيس** من جهه على التوازن الضرورى فى توزيع القوات

And **the Presidency** said **President** emphasized on the necessary balance of forces distribution

This is an example of a sentence where the CRF miss-tag the opinion holder in an indirect speech event subjective statement (Type 1 Holder) by tagging (the presidency organization الرئاسة) instead of (the president الرئيس), the true opinion holder.

**Pattern Experiment Correct Examples:**

- واكد **النائب مسلم البراك** انه لا يرى ضرورة الربط بين اسقاط ديون العراق وقروض المواطنين

And **the parliament member Muslim Al-Barrak** confirmed he doesn't see any need for binding Iraq's debt payment to people loans

**False Positive in Pattern Experiment:**

- وقال **المال العام** يمكن ان يحقق حاليا العدالة التي يتحدثون عنها

And said **the public finance** can achieve whatever justice they are speaking about

**False Negative in Pattern Experiment:**

- واستهجن **عكاش** نشر هذا الكتاب وفيه هذا الطعن المرفوض جملة وتفصيلا

And **Okash** condemned publishing this book containing such unacceptable defamation

# 7. CONCLUSIONS AND FUTURE WORK

This paper presents a leading research work in Arabic opinion holder field. Opinion source identification in Arabic language are explored using two approaches, we conclude that sequential tagging ML classifiers outperform patterns in terms of recall and precision. Moreover, patterns do not contribute significantly to the results after they are encoded as a CRF feature. NE and Semantic field features are crucial for opinion source detection and they partially compensate the lack of parser features. We are going to explore the possibility to enhance the Arabic Opinion Holder Extraction task performance while utilizing a robust Arabic lexical or dependency parser constituents.

## ACKNOWLEDGEMENTS

All authors thank profoundly Engineers Amr Magdy and Marwa Magdy, NLP researchers, for their coding and translation support through the development of this paper

**Mohamed Elarnaoty** is a Teaching Assistant at Computer Science Department, Faculty of Computers and Information, Cairo University. He is interested in Mathematics, Machine Learning, Information Extraction, Opinion Mining, and Natural Language Processing fields.

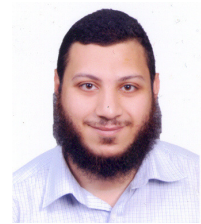

**Samir AbdelRahman** is an Associate Professor at Computer Science Department, Faculty of Computers and Information, Cairo University. His M.Sc. and Ph.D. had been received from Cairo University in Computer Science Specialty. His main research interests include Machine Learning, Text Mining, NLP and Agent Negotatians fields. He published over 40 publications and supervised over 80 theses and graduation projects. He was awarded the best faculty professr for three consequestive years (2005-2008). Since then, he has been an NLP visiting researcher in three USA world ranked universities: Univeriveriy of Minnesota, Vanderbilt University and University of Illinois,Urbana-Champaign.

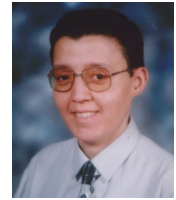

**Dr. Aly Aly Fahmy**, is the former Dean of the Faculty of Computing and Information, Cairo University and a Professor of Artificial Intelligence and Machine Learning. He was the Director of the first Center of Excellence in Egypt in the field of Data Mining and Computer Modeling (DMCM) in the period of 2005-2010. DMCM was a virtual research center with more than 40 researchers from universities and industry. He is currently involved in two main activities. The implementation of Cairo University theses mining project to assist in the formulation of the University strategic research plan for the coming 2011 – 2015 and the advancement of the Arabic Language technologies.

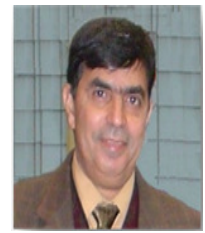

**Dr. Fahmy's** main research areas are: Data and Text Mining, Computational Linguistics, Text Understanding and Automatic Essay Scoring and Technologies of Man- Machine Interface in Arabic.